\documentclass[12pt]{article}
\usepackage{psfig}

\advance\textheight by\topskip

\def\arcsec{\hbox{$^{\prime\prime}$}}
\def\fm{\hbox{$.\!\!^{\rm m}$}}
\def\farcs{\hbox{$.\!\!^{\prime\prime}$}}

\begin{document}

\title{\bf
RC J1148+0455
IDENTIFICATION: GRAVITATIONAL LENS OR GROUP OF GALAXIES ?
}

\author{
Oleg~V.~Verkhodanov,~Alexander~I.~Kopylov,~Yurij~N.~Parijskij,
\and Natalia~S.~Soboleva,~Olga~P.~Zhelenkova,~Adelina~V.~Temirova,
\and{\it Special Astrophysical Observatory of Russian Academy of
Sciences}
\and{\it         357147 Karachai-Cherkesia, N.Arkhyz, Russia
}
\and{\it E-mail: vo@sao.ru, akop@sao.ru, par@sao.ru, sns@fsao.spb.su}
\and {\it zhe@sao.ru,~tem@fsao.spb.su }
\and Joshua~Winn,~Andre~Fletcher,~Bernard~Burke
\and {\it Massachusetts Institute of Technology, USA}
\and {\it E-mail: jnwinn@mit.edu, fletcher@mit.edu, bfburke@mit.edu}
}

\date{}
\maketitle

\begin{abstract}
The structure of the radio source RC~B1146+052 of the ``Cold''catalogue
is investigated by data of the MIT--GB--VLA survey at 4850~MHz.
This source belongs to the steep spectrum radio sources subsample of
the RC catalogue. Its spectral index is $\alpha$ = -1.04.
The optical image of this source obtained with 6m telescope is analysed.
The possible explanations of the complex structure of radio components
are considered.
\end{abstract}

\section {Introduction}

Using RATAN--600 ``Cold'' catalog (Parijskij et al., 1996),
we have selected
steep spectrum sources having two-component structure of the FRII type
(Fanaroff and Riley, 1974).
Among the ``Cold'' catalog
we have mapped with VLA
about 100 such objects. The majority of these objects is identified
with elliptical galaxies up to 24--25$^m$~$R$
(Kopylov et al., 1995; Parijskij et al., 1998), which were observed
with the 6m telescope of the Special astrophysical observatory
by the program ``Big Trio''.

To understand a structure of the object  we have used VLA observations
from the MIT--GB survey (Bennet et al., 1986), which crosses
the ``Cold'' survey region.  The detailed maps have been obtained
for 69 objects of RC catalog using MIT--GB--VLA data
(Fletcher, 1996).

In this paper we study a nature of the radio source
RC~B1146+052 (RC~J1148+0455) of the ``Cold'' catalog using
radio maps obtained in the MIT--GB--VLA survey showing a complex
structure of components and 6m telescope data.

There were several explanations of the
complex structure of radio components. One of them describes
the Eastern components as a possible candidate to the gravitational lens
(Fletcher, 1998).
We have studied this problems in details.

\section {Radio data}

\begin{figure}
\centerline{\psfig{figure=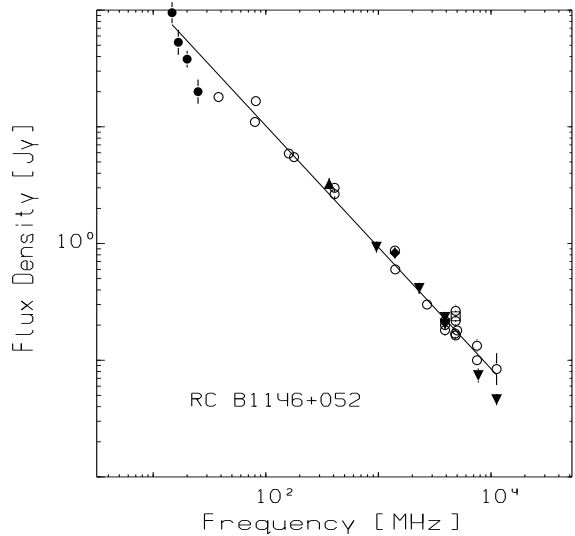,width=10cm,angle=-90}}
\caption{Spectrum of radio source RC~B1146+052 including
UTR data (Braude et al., 1979), 4C data (Pilkington \& Scott, 1965:
object 4C+05.53) and Parks data (Otrupcek \& Wright, 1991). Spectrum
has been prepared using data of the CATS database.
}
\end{figure}

Using the CATS database (Verkhodanov et al., 1998) we have prepared
a radio spectrum of the source.
The spectral index $\alpha$ obtained with linear fitting
by the least square method equals to
 --1.04 ($S \sim \nu^{\alpha}$). This allows us to classify this object
as the ultra steep spectrum source.

To study a radio structure of RC~B1146+052 we have used
the VLA survey data
archived in the Massachusetts Institute of Technology.
A resolution of the image is 0\farcs5 at
4850 MHz in A-configuration.

\begin{figure}
\centerline{\psfig{figure=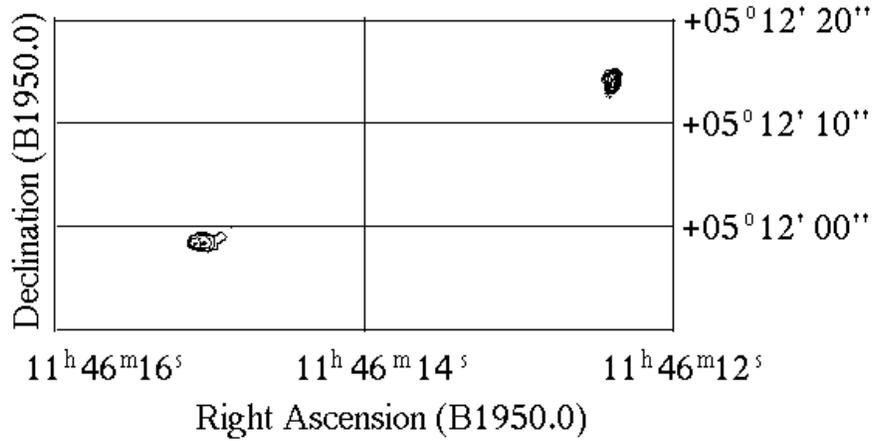,width=12cm,angle=-90}}
\caption{Two component structure of the object RC~B1146+052.
A component á (Western) is on the right,
a component B (Eastern) is on the left}
\end{figure}

\begin{figure}
\centerline{
\hbox{
\psfig{figure=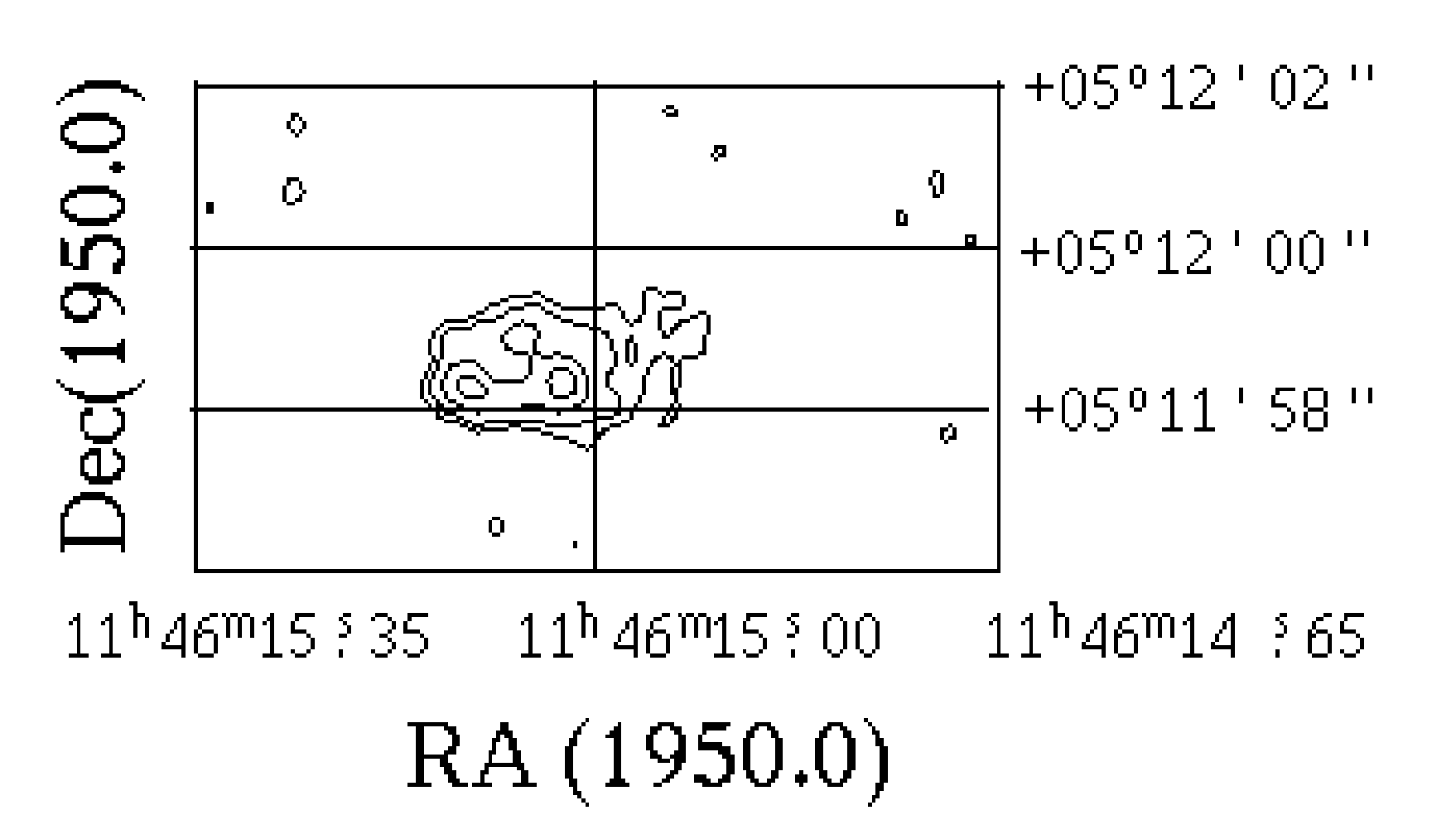,width=8cm,angle=-90}
\psfig{figure=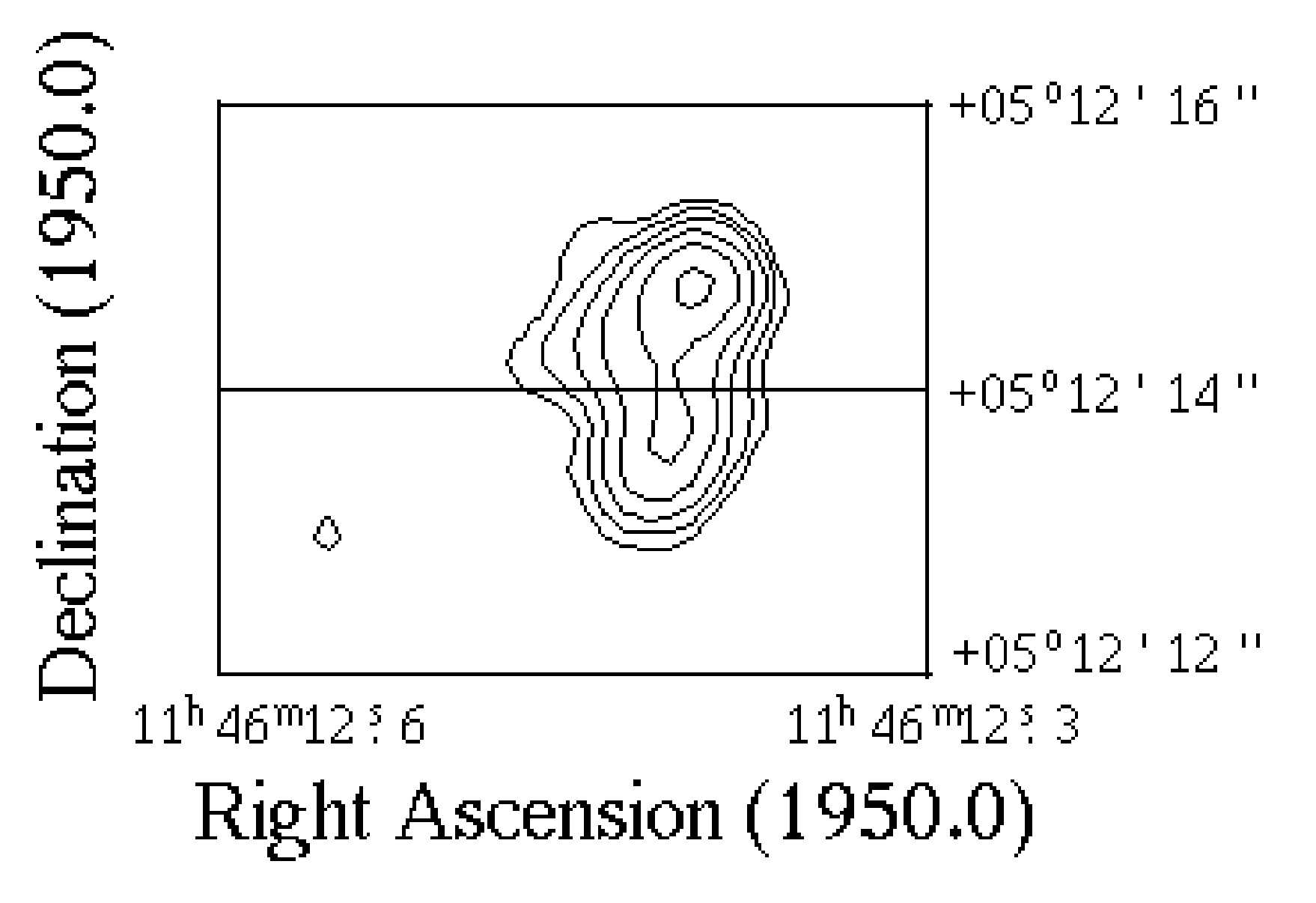,width=8cm,angle=-90}
}}
\caption{A components structure (from right to left: A and B)
of RC~B1146+052.}
\end{figure}

\begin{table}
\caption{Positions and flux densities at 4850 MHz of components of
RC~B1146+052, by the MIT-GB-VLA survey data}
\begin{center}
\begin{tabular}{|lccrr|}
\hline
radio&$\alpha+\delta(B1950.0)$&$\alpha+\delta(J2000.0)$&$S_{peak}$&$S_{int}$\\
component&                    &                        & mJy/beam & mJy \\
\hline
A        & 114612.38+051214.8 &  114846.48+045534.2 & 28.4 & 166    \\
B        & 114615.07+051158.6 &  114849.16+045517.9 &      &  96    \\
B2       & 114615.03+051158.4 &  114849.12+045517.8 &  5.7 &        \\
B1       & 114615.06+051158.9 &  114849.16+045518.3 &  4.3 &        \\
B3       & 114615.10+051158.3 &  114849.19+045517.7 &  6.2 &        \\
\hline
A+B data & 114613.37+051207.6 &  114847.46+045527.0 &      & 263    \\
\hline
\end{tabular}
\end{center}
\end{table}

The component A (Fig.3 and 4) has a curved structure and been classified
earlier (Lawrence et al., 1986; Parijskij et al., 1996)
as a double independent radio source.
The component B has a complex structure with 3 maxima.
If this component is a separate source and
taking into account its complex structure it is possible to
consider it a candidate to the gravitational lensed source
(Fletcher, 1998). The coordinates of the both component maxima
at 4850 MHz are given in Table 1.
The distance between components is 43\arcsec.

\section {Optical identification}

In Fig.4 are plotted contours of the radio source RC~B1146+052
by the MIT--GB--VLA survey and NVSS survey overlaid
on the 6m telescope image in R--filter.

\begin{figure}
\centerline{\psfig{figure=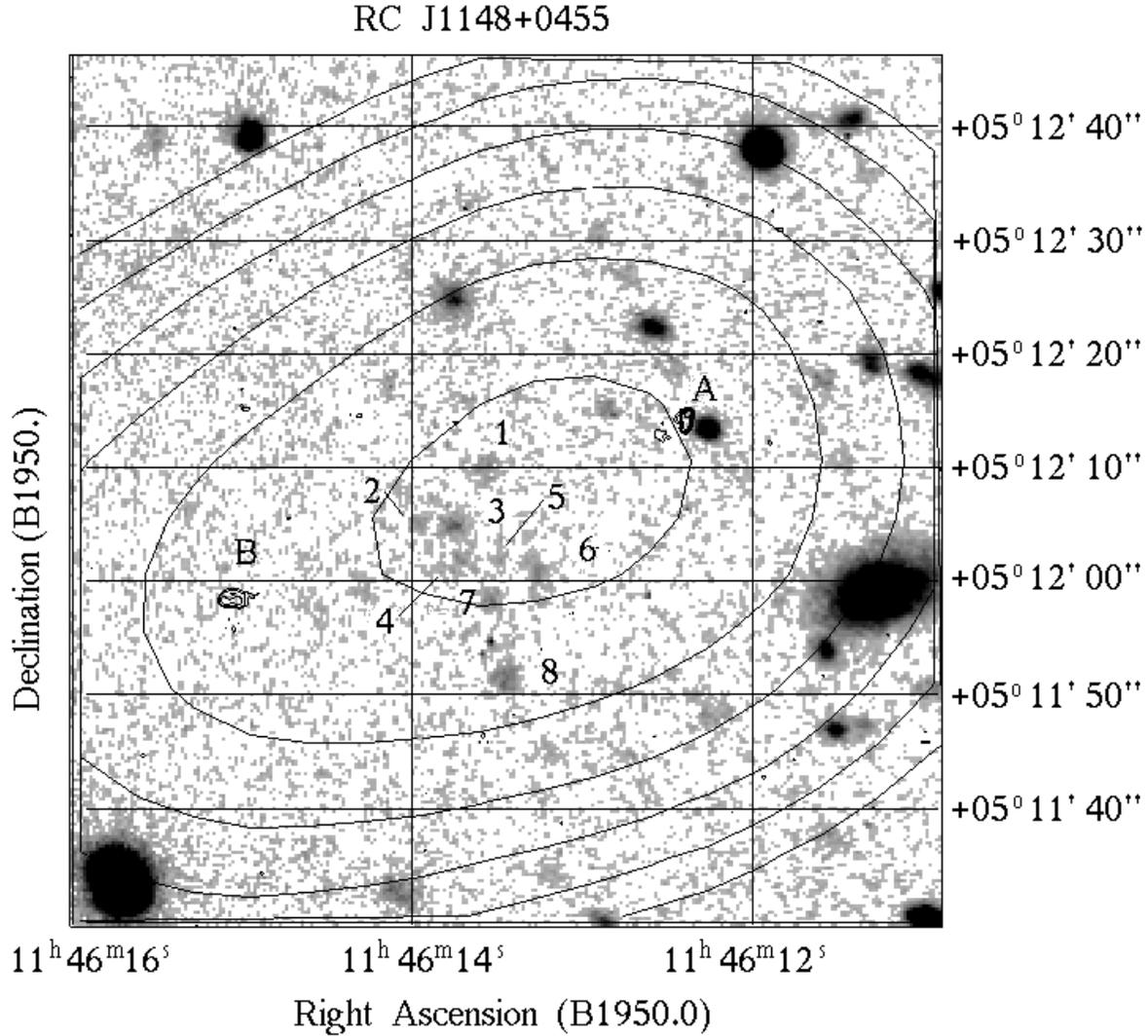,width=16cm}}
\caption{
MIT--GB--VLA survey contours (A and B) of RC~B1146+052 at 4885 MHz
(from the level of 1 mJy with factor 2) and NVSS contours at 1400 MHz
(from the level of 80 mJy with factor 2) overlaid on the 6m telescope
R--image.
}
\end{figure}

A group of 8 galaxies lies near the center of the NVSS source.
The brightness one of them lies on an axis connecting both components.
It has a magnitude of 23\fm2$\pm$0\fm2 in the R--filter within
an aperture of 3''.
Some characteristics of this group galaxies are given in Table 3.

\begin{table}
\caption{Coordinates and magnitude of galaxies in the group}
\begin{center}
\begin{tabular}{|cccl|}
\hline
 Number  &$\alpha(B1950.0)$&$\delta(B1950.0)$&  R-mag    \\
\hline
%  1         & 11 46 13.58  & +5 12 10.2  & -2.957  \\
%  2         & 11 46 13.96  & +5 12 05.3  & -2.273  \\
%  3         & 11 46 13.77  & +5 12 04.9  & -3.403  \\
%  4         & 11 46 13.73  & +5 12 01.9  & -2.934  \\
%% 5         & 11 46 13.54  & +5 12 01.8  & -2.232  \\
%  5         & 11 46 13.48  & +5 12 01.9  & -2.232  \\
%  6         & 11 46 13.27  & +5 12 00.1  & -2.935  \\
%  7         & 11 46 13.55  & +5 11 58.3  & -2.825  \\
%  8         & 11 46 13.45  & +5 11 51.7  & -3.562  \\
%
  1         & 11 46 13.58  & +5 12 10.2  & 23.8  \\
  2         & 11 46 13.96  & +5 12 05.3  & 24.3  \\
  3         & 11 46 13.77  & +5 12 04.9  & 23.2  \\
  4         & 11 46 13.73  & +5 12 01.9  & 23.7  \\
  5         & 11 46 13.48  & +5 12 01.9  & 24.4  \\
  6         & 11 46 13.27  & +5 12 00.1  & 23.7  \\
  7         & 11 46 13.55  & +5 11 58.3  & 23.6  \\
  8         & 11 46 13.45  & +5 11 51.7  & 23.1  \\
\hline
\end{tabular}
\end{center}
\end{table}

\section{Discussion}

Using all the data discussed above we conclude that
the radio source RC~B1146+052 can not be a result of two blending objects,
but it has a type of FRII. In this case, the parent galaxy, apparently,
belongs to the group of galaxies.
The curved form of the A component can be explained as a result of
interaction with the gaseous environment, and a form of the B component
can be explained in the frames of the jet rotation hypothesis or
also as a result of interaction with the gaseous environment
as in Cygnus~A.

{\small\rm
{\bf Acknowledgements}
This work has been supported in part by Russian Foundation of Basic
Research (grants NN 99-07-90334 and 99-02-17114), and by the Federal Programs
``Astronomy'' (grants 1.2.1.2 and 1.2.2.4), and ``Integration''
(grant No 578).

}

\end{document}